\PassOptionsToPackage{table}{xcolor}

\documentclass[sigconf]{acmart}

\AtBeginDocument{%
  \providecommand\BibTeX{{%
    \normalfont B\kern-0.5em{\scshape i\kern-0.25em b}\kern-0.8em\TeX}}}

\setcopyright{acmcopyright}
\copyrightyear{2018}
\acmYear{2018}
\acmDOI{10.1145/1122445.1122456}

\acmConference[PACT '22]{PACT '22:  International Conference on
Parallel Architectures and Compilation Techniques (PACT)}{October 10--12, 2012}{Chicago,IL}
\acmBooktitle{PACT '22: International Conference on
Parallel Architectures and Compilation Techniques (PACT), October 10--12, 2018, Chicago, IL}
\acmPrice{15.00}
\acmISBN{978-1-4503-XXXX-X/18/06}

\usepackage{array}

\usepackage{algorithm}
\usepackage{algpseudocode}
\algnewcommand\algorithmicforeach{\textbf{for each}}
\algdef{S}[FOR]{ForEach}[1]{\algorithmicforeach\ #1\ \algorithmicdo}
\usepackage{tablefootnote}
\usepackage[inline]{enumitem}
\usepackage[normalem]{ulem}
\setlist[enumerate]{itemsep=0mm}

\newcommand{\TabHead}{\bfseries\small}

\newcommand{\add}[1]{\textcolor{black}{#1}}
\newcommand{\addd}[1]{\textcolor{black}{#1}}

\definecolor{darkgrn}{rgb}{0, 0.75, 0}

\usepackage{xargs}                      
\usepackage[colorinlistoftodos,prependcaption,textsize=footnotesize]{todonotes}
\usepackage{xspace}

\newcommandx{\ak}[2][1=]{\todo[color=green!50,#1]{\sf \textbf{Ananth:} #2}\xspace}
\newcommandx{\MH}[2][1=]{\todo[color=blue!50,#1]{\sf \textbf{MH:} #2}\xspace}
\newcommandx{\mm}[2][1=]{\todo[color=blue!50,#1]{\sf \textbf{M\&M:} #2}\xspace}

\newcommandx{\dt}[2][1=]{\todo[color=green!50,#1]{\sf \textbf{Dingwen:} #2}\xspace}

\newcommand{\huffmax}{\texttt{Huffmax}}
\newcommand{\bitmax}{\texttt{Bitmax}}
\newcommand{\hbmax}{\texttt{HBMax}}
\newcommand{\ripples}{\texttt{Ripples}}

\begin{document}

\title{HBMax: Optimizing Memory Efficiency for Parallel Influence Maximization on Multicore Architectures}


\settopmatter{authorsperrow=3}

\newcommand{\AFFIL}[4]{%
    \affiliation{%
        \institution{\small #1}
        \city{#2}\state{#3}\country{#4}
    }
    }
    
\author{Xinyu Chen}{\AFFIL{Washington State University}{Pullman}{WA}{USA}}
\email{xinyu.chen1@wsu.edu}

\author{Marco Minutoli}{\AFFIL{Pacific Northwest National Laboratory}{Richland}{WA}{USA}}
\email{marco.minutoli@pnnl.gov}

\author{Jiannan Tian}{\AFFIL{Washington State University}{Pullman}{WA}{USA}}
\email{jiannan.tian@wsu.edu}

\author{Mahantesh Halappanavar}{\AFFIL{Pacific Northwest National Laboratory}{Richland}{WA}{USA}}
\email{mahantesh.halappanavar@pnnl.gov}

\author{Ananth Kalyanaraman}{\AFFIL{Washington State University}{Pullman}{WA}{USA}}
\email{ananth@wsu.edu}

\author{Dingwen Tao}{\AFFIL{Washington State University}{Pullman}{WA}{USA}}
\authornote{Corresponding author.}
\email{dingwen.tao@wsu.edu}


\settopmatter{printacmref=false} 
\settopmatter{printfolios=true}
\renewcommand\footnotetextcopyrightpermission[1]{} 

\begin{abstract}
  Influence maximization aims to select $k$ most-influential vertices or seeds in a network, where influence is defined by a given diffusion process. 
  Although computing optimal seed set is NP-Hard, efficient approximation algorithms exist. 
  However, even state-of-the-art parallel implementations are limited by a sampling step that incurs large memory footprints. This in turn limits the problem size reach and approximation quality. 
  In this work, we study the memory footprint of the sampling process collecting reverse reachability information in the IMM \add{(Influence Maximization via Martingales)} algorithm over large real-world social networks. We present a memory-efficient optimization approach (called \hbmax) based on \add{Ripples}, a state-of-the-art multi-threaded parallel influence maximization solution. 
  Our approach, \hbmax, uses a portion of the reverse reachable (RR) sets collected by the algorithm to learn the characteristics of the graph. Then, it compresses the intermediate reverse reachability information with Huffman coding \add{or bitmap coding}, and queries on the \add{partially decoded data, or directly on the} compressed data to preserve the memory savings obtained through compression.  
  Considering a NUMA architecture, we scale up our solution on 64 CPU cores and reduce the memory footprint by up to $82.1\%$ with average $6.3\%$ speedup \addd{(encoding overhead is offset by performance gain from memory reduction)} without loss of accuracy. \add{For the largest tested graph Twitter7 (with 1.4 billion edges), {\hbmax} achieves $5.9\times$ compression ratio and $2.2\times$ speedup.}
\end{abstract}




\maketitle

\section{Introduction}\label{sec:Introduction}
A graph, $G=(V,E)$, captures complex relationships between a set of entities represented as nodes or vertices ($V$), through binary relations expressed as edges or links ($E$). Graph analytics provides a set of algorithms such as centrality measures, community detection, shortest paths, and network flow to enable decision making on data presented as graphs. The 
ubiquity of massive data from domains such as social networks and life sciences has enabled application of graph analytics on numerous domains with varying degrees of scale. 
A fundamental limitation to the application of graph analytics is the massive memory requirement of algorithms. Since many graph algorithms have irregular accesses to memory and low (arithmetic) computation, the performance of memory system becomes critical.   

Given a directed graph $G=(V,E,\omega)$, where $\omega$ represents edge weights corresponding to the influence of node $x$ on node $y$ for an edge $(x,y)$; a diffusion model, and a budget $k$; the {\em Influence Maximization} (IM) problem is an optimization problem to identify a set of $k$ vertices which when activated  result in a maximal number of expected activation in $G$, where activation is defined for a given model of diffusion. \add{The IM problem came from} viral marketing, \add{where a company tried to create a cascade of product adoption through the word-of-mouth effect by choosing a set of influential individuals and giving them free samples of the product.} \add{Now It} has numerous applications \cite{kempe2003maximizing, leskovec2007dynamics, minutoli2020preempt} in domains such as politics, public health, bioinformatics,  and sensor networks. 

\begin{figure}[t]
  \centering
  \includegraphics[width=.9\linewidth, trim={0 .2in 0 .15in}]
  {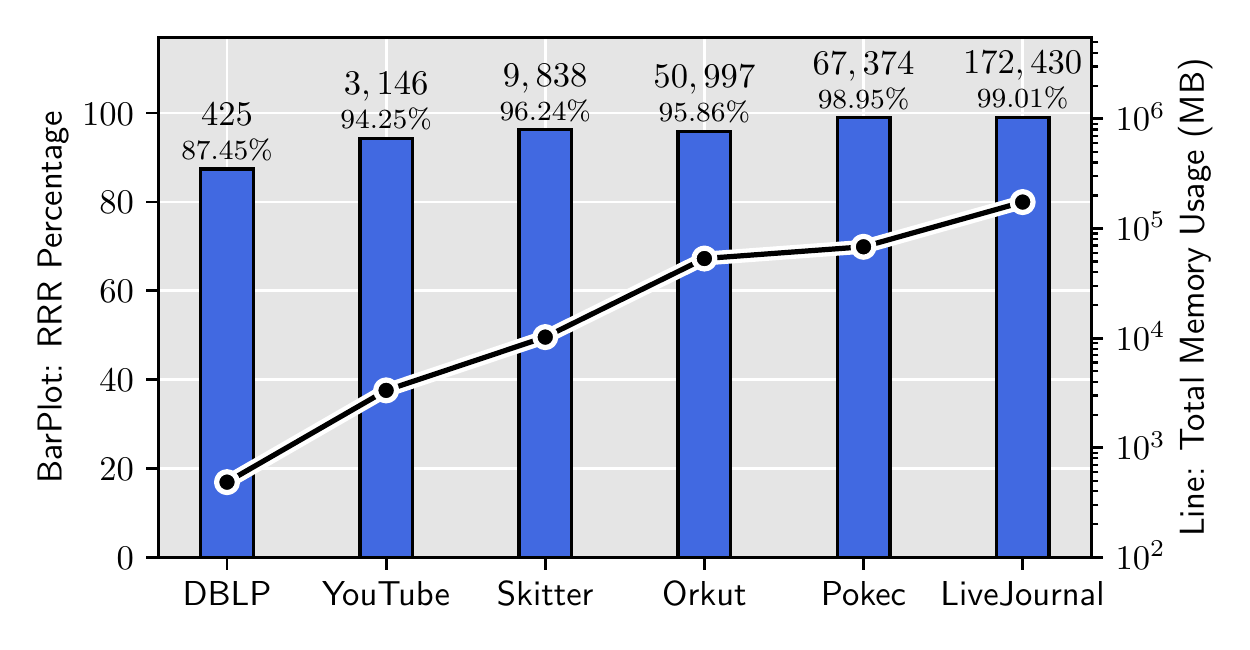}%
  \caption{Memory usage breakdown of Ripples \cite{minutoli2019fast}, where the blue portions represent the computation of random reverse reachable (RRR) paths (87 to 99\% of memory usage). 
}
  \label{fig:mem_usage} 
\end{figure}

Finding the top $k$ influential vertices (i.e., seeds) in a graph can be formulated as a discrete optimization problem that has been shown to be NP-Hard  \cite{kempe2003maximizing}.
Consequently, a few key efficient approximation solutions have been developed.  
\textcolor{black}{For instance, Kempe {\it et al.} \cite{kempe2003maximizing} attempts to find an approximate solution by hill-climbing on a large number of Monte Carlo (MC) diffusion processes (typically around $10^4$).  Borgs {\it et al.} \cite{borgs2014maximizing} greatly improves the efficiency of MC simulations by exploiting the potential overlap in the vertex space among multiple simulated paths of diffusion. Their approach enumerates random reverse reachable (RRR) sets and using them to select influential vertices that occur frequently in them. 
Tang {\it et al.}  \cite{tang2015influence} further extend this work with a two-phase IMM algorithm (including sample estimation and seed selection) to improve the determination of the sampling effort through a martingale strategy (will be discussed in detail in Section \ref{sec:Background}).}
The targeted quality \add{is a $(1-1/e-\epsilon)$-}approximation. \add{The error factor} $\epsilon$ plays an important role in the overall performance of the IMM algorithm. \add{Smaller $\epsilon$ will produce} a more accurate approximation \add{by making} the algorithm \add{carry out} more MC trials. With the growing size of social networks and the goal of getting high-quality approximation, the IMM algorithm is both computationally challenging and memory demanding.
To alleviate this challenge, parallel computing has been introduced to reduce the execution time of the IMM algorithm.
Ripples, proposed by Minutoli {\it et al.} \cite{minutoli2019fast} is the state-of-the-art parallel implementation of the IMM algorithm. It can solve the time-cost challenge by parallelizing the computation and scaling up in both the shared-memory and distributed-memory architectures. 

However, few studies in the literature address the memory-intensive challenge, which should have a standalone place in the research on the IMM algorithm. More specifically, the reason to address the memory challenge of the IMM algorithm is two-fold: (1) Although the aggregated memory capacity of today's high-performance computing (HPC) systems is ever-increasing, such distributed computing resource is not readily available to most. Thus, reducing the memory usage on shared-memory systems helps to solve larger influence maximization problems with limited computational resources. \textcolor{black}{(2) There is a vast difference between the input graph size and the memory footprint during the computation \cite{ghosh2021single}. This memory inflation (by the algorithm) is particularly pronounced for a stochastic graph application such as IMM. For instance, for the six graphs studied in Figure~\ref{fig:mem_usage}, the ratio of the peak memory usage to the input size varies from 30$\times$ to 165$\times$---implying  a one to two orders of magnitude increase in memory requirement during the course of computation to store the intermediate results. }

To this end, \add{we propose a memory-efficient parallel influence \underline{max}imization algorithm using \underline{H}uffman coding and \underline{B}itmap coding (called \hbmax) and implement it based on the state-of-the-art implementation Ripples \cite{minutoli2020curipples}}. Specifically, by profiling the memory footprint of Ripples on large real-world graphs, we identify the most demanding portion of the algorithm and different demands with different types of graphs. With these characteristics, we propose a block-based workflow that leverages the Huffman coding or \add{bitmap coding} to save the intermediate MC simulation results in a compressed format. \add{The subsequent analysis can be performed on partially decompressed Huffman encoded data or directly performed on the bitmap encoded data: neither of the two schemes needs to fully decompress the data to the original size,} so as to preserve the memory savings. 
\add{To the best of our knowledge, this work represents the most space-efficient parallel IMM on very large graphs.} Our main contributions are summarized as follows.
\begin{enumerate}[leftmargin=*,noitemsep,topsep=1pt]
    \item We conduct a comprehensive characterization and profiling of different real-world graphs to understand the impact of their features on the memory footprint based on the state-of-the-art solution for the influence maximization problem.
    \item We \add{identify the intermediate RRR sets sampled from MC simulations vary from skewed-distributed to flat-head distributed. We} propose a scalable ``compress-to-compute'' based IMM method (called {\hbmax}) that leverages the Huffman coding or bitmap coding to reduce the memory footprint of saving intermediate RRR sets \add{based on these characteristics}.
    \item \add{We propose two efficient seed selection approaches based on the two encoding methods. Specifically, for Huffman-encoded intermediate RRRs, we exploit data locality to query the compressed RRRs without fully decoding them; for bitmap-encoded RRRs, we directly query the compressed RRRs with bit operations. Moreover, we propose a new parallel max-reduction method for finding the vertex with the maximum frequency \addd{to improve the scalability of seed selection.}
    }
    \item We evaluate our \hbmax~ and compare it with the original Ripples implementation on eight large graphs (\add{Ripples cannot run the largest two graphs on the tested machine with 376 GB RAM).} Experiments show that \hbmax~ reduces the memory usage by up to \add{$82.1\%$ with $6.3\%$} speedup (the encoding overhead is offset by performance gain from memory reduction). Moreover, \hbmax~ can reduce the overall time by up to $79\%$ with the same memory footprint when compared to Ripples. 
\end{enumerate}

The rest of this paper is organized as follows. In Section \ref{sec:Background}, we present background information about the IM problem, state-of-the-art algorithms and implementations, Huffman coding, bitmap coding and related work. 
In Section~\ref{sec:Profiling}, we profile the memory footprint of Ripples with different graphs and characterize input graphs into \textcolor{black}{two} main categories.  
In Section~\ref{sec:Method}, we propose our three optimizations for memory footprint reduction and performance/scalability enhancement. 
In Section~\ref{sec:Evaluation}, we evaluate our optimized IMM solution and compare it with the state-of-the-art method on \textcolor{black}{large} graphs. 
In Section~\ref{sec:Conclusion}, we conclude our work and discuss the future work.

\section{Background and Related Work}\label{sec:Background}
In this section, we introduce the influence maximization problem, the RIS and IMM algorithms, Ripples software, Huffman/bitmap compression, and compressed-based graph analytics.

\subsection{Influence Maximization Problem}
Let $G=(V, E)$ be a graph with $n$ vertices and $m$ edges. Given $G$ and a stochastic diffusion process, the influence maximization problem is one of identifying a set $S$ of top $k$ vertices in $G$ (called ``seeds'') that maximizes the expected influence spread ($\mathbb{E}[I(S)]$) in $G$, measured by the number of activated vertices. Kempe {\it et al.} \citep{kempe2003maximizing} have shown that $\mathbb{E}[I(S)]$ is a submodular function of $S$ for two simple but powerful diffusion models: the independent cascade model (IC) and the linear threshold model (LT). Submodular functions have the property of diminishing marginal gains and leads to efficient approximation algorithms \cite{nemhauser1978analysis}. Leveraging the submodularity framework, they propose a greedy hill-climbing algorithm that offers an approximation guarantee of $1-1/e-\epsilon$ \cite{kempe2003maximizing}.

The IC and LT diffusion models are broadly studied in the literature. The IC model comes from the physics of interacting particles. In this model, each newly activated vertex has a single chance to activate its neighbors. Assuming a directed graph $G=(V,E,\omega)$ as an example, each edge $(u,v) \in E$ is assigned with a probability $p(u,v)$ to trigger the activation of $v$ from $u$. When the vertex $u$ gets activated at time $t$, then at time $t+1$, the vertex $v$ gets activated with the probability $p(u,v)$. 

In contrast, the LT model captures mass behaviours. Each vertex has a threshold modeling their resistance to adopting the mass behavior and each edge $(u,v)$ has a weight $w$ representing the capacity of $u$ to influence $v$. At time $t$, a vertex $v$ gets activated if the sum of the weights on the incoming edges from its already active neighbors exceeds its threshold. \add{It is worth noticing that the edge weights $\omega$ are provided by input graphs, but they are not corresponding to the probability of activation. Thus they only affect the LT model instead of the IC model.}

The LT diffusion model tends to produce very small RRR sets and more than $99\%$ of them end up with less than 10 vertices \cite{minutoli2020curipples}. On the contrary, the sizes of RRR sets from the IC model is less skewed and  tend to be of larger size. 
From a utility standpoint, the IC model is more generally applicable to abstract a range of diffusion processes. \add{It can be viewed as a special case of the} Susceptible-Infected-Recovered (SIR) models \cite{minutoli2020preempt} used in epidemics. However, its large memory footprint (as shown in Figure~\ref{fig:mem_usage}) limits its scalability, so we focus on the IC diffusion model in this paper.


\subsection{RIS, IMM, and Ripples}
\noindentparagraph{\textbf{Reverse Influence Sampling}.}  Following the seminal work  \cite{kempe2003maximizing}, \citet{borgs2014maximizing}  proposed an approximation algorithm based on the idea of Reverse Influence Sampling (RIS). Their scheme attempts to find which are the most likely causes of activation for each vertex in the graph. The algorithm randomly samples vertices $v$ and simulates the diffusion model in reverse collecting sets of vertices which may cause the activation of $v$. The seed sets $S$ is later decided by solving a maximum coverage problem over the sets collected through RIS. This approach provides the same approximation guarantee as the greedy hill-climbing algorithm \cite{kempe2003maximizing}.

The RIS approach is the fundamental building block of the IMM algorithm from \citet{tang2015influence}, and its state-of-the-art parallel implementation Ripples \cite{minutoli2019fast} is the starting point for our work. In the following, we give important definitions and a high-level description of the fundamental building blocks of the IMM and Ripples algorithm. We direct the reader to the original work for a more exhaustive presentation. 

\vspace{-2mm}
\begin{definition} [Reverse Reachable (RR) Set]
  Let $\hat{G}=(V,\hat{E})$ be the transposed graph obtained from $G=(V, E)$ by inverting the orientation of all the edges in $E$. The reverse reachable set of a vertex $v$ is the set of vertices $u \in V$ that are reachable from $v$ in $\hat{G}$.
\end{definition}

\vspace{-3mm}
\begin{definition} [Random Reverse Reachable (RRR) Set]
  Let $g=(V, E')$ be a subgraph of $G=(V, E)$ obtained by edge removal by retaining only the active edges during a simulation of a diffusion process $M$ on $G$. A random reverse reachable set $RR_g(v)$ for a vertex $v$ is the set of vertices $u \in V$ that are reachable from $v$ in $\hat{g}$, where $\hat{g}$ is the transposed graph obtained from $g$.
\end{definition}

\vspace{-2mm}
In the following discussion, we use \textit{RRR sets} to refer to the collection of random reverse reachable sets. We use \textit{rr} or \textit{sample} to refer to one RRR. Since each RRR is a set of vertices, we refer the cardinality of an RRR as the size of RRR. 

\vspace{-1mm}
\noindentparagraph{\textbf{IMM and Ripples}.} \add{Although the reverse reachable scheme of \citeauthor{borgs2014maximizing} greatly improves the efficiency of MC simulations, it could overestimate the required number of MC simulations and waste a lot of computation.} The IMM algorithm proposed by \citet{tang2015influence} adopts a two-phase design to algorithmically determine the sampling effort by leveraging a martingale strategy\footnote{\add{The martingale strategy is a betting strategy in which a person doubles the bet every time they lose knowing that they must win at some point. It first samples a small number of RRR sets to calculate the achieved influence and then doubles the number of samples until the influence reaches an estimated lower bound. }}. \add{This strategy} greatly improves the \add{performance so that IMM} algorithm can analyze large graphs with millions of vertices. 

Specifically, the two-phase design works as follows.
In the \emph{Sampling} phase, the algorithm will produce $\theta$ \textit{RRR}s starting from random vertices in the input graph and simulating the diffusion model (IC or LT) in reverse. In the case of the IC and LT model, the task of generating a RRR set resembles a randomized breadth-first search (BFS) where only a subset of the neighbors of a vertex enter the next frontier.
To estimate the required sampling effort ($\theta$) to achieve the approximation guarantee (controlled through the parameter $\epsilon$), the algorithm uses two important results. \citeauthor{borgs2014maximizing} showed that the fraction of RRR covered during the seed selection process is an unbiased estimator of the influence function while \citeauthor{tang2015influence} were able to prove a lower bound on the sampling effort using an estimation of the influence function. In Equation (\ref{eq:theta}), \citeauthor{tang2015influence} shows the algorithm starts with a guess on $\theta$ and doubles the sampling effort at every iteration until the exit condition derived from the lower lower bound is satisfied and the final value of $\theta$ is computed. \add{Here $n$ is the number of vertices, $k$ is the number of seeds, $\epsilon$ is the error factor, and $\theta$ is the required number of sampling. The larger $n$ and $k$ are, the larger $\theta$ will be. On the other hand, small error factor $\epsilon$ will increase $\theta$ non-linearly. This process resembles to the martingale betting strategy. } 
\begin{equation}\label{eq:theta}
    \theta_i = \frac{(2+\frac{2\sqrt{2}}{3}\epsilon)\cdot(\log{n \choose k}\cdot \log(n)+\log\log_2(n)) \cdot 2^i}{2\epsilon^2} 
\end{equation}  

The \emph{Seed Selection} algorithm is based on the greedy maximum coverage algorithm \cite{vazirani2001approximation}. The method iterates $k$ times over the RRR sets to select the vertex $v$ appearing most frequently. At every iteration $i$, the RRR sets covered by one of the $i$ seeds already selected are ignored. \citet{minutoli2019fast,minutoli2020curipples} devise efficient parallel schemes that perform the counting by either updating the state of the previous iteration or recount from scratch when more profitable.

\add{Ripples \cite{ripples} is a state-of-the-art parallel software framework that provides fast and scalable implementation for the IM problem. According to \cite{minutoli2019fast, minutoli2020curipples}, its CPU version provides a speedup of 580$\times$ over the best sequential baseline using 1024 nodes, and its CPU+GPU version provides a speedup of 760$\times$ over the best sequential baseline.} 

\vspace{-2mm}

\subsection{Huffman Coding and Bitmap Compression}
\textit{Huffman coding} is a classical data compression technique \cite{huffman1952method}. 
It assigns variable-length codes to encode target characters based on their relative frequency, which gets better reduction when the data has skewed distribution \cite{roth1993database}. The Huffman codes are prefix-free and are typically created as a binary tree with the encoded characters stored at the leaves. 
There are works that adopt Huffman coding to reduce the memory footprint. For example, 
Suontausta and Tian \cite{suontausta2003low} applied Huffman coding for efficiently storing decision tree parameters to minimize the memory footprint. 
Ficara {\it et al.} \cite{ficara2010enhancing} adopted Huffman coding to improve counting bloom filters in terms of fast access and limited memory consumption.

\add{\textit{Bitmap} is a mapping from some domain (e.g., a range of integers) to bits \cite{foley1996computer}. 
It not only reduces the data size but also allows efficient direct operations such as binary logic operations \cite{martinez2012efficient}.
Thus, many efficient bitmap compression schemes have been extensively studied in the
database systems, such as BBC, WAH, EWAH, and Roaring \cite{wang2017experimental}.}
However, none of these work exploited data analytics on the Huffman or bitmap encoded data directly.

\subsection{Graph Compression and Data Analytics on Compressed Data}
Due to the growing sizes of graphs, researchers have been studying compression techniques for graphs.
For example, Randall {\it et al.} \cite{randall2002link} tried to compress the links of web graphs by leveraging the locality of web graphs. 
Ligra \cite{shun2013ligra} and SlimGraph \cite{besta2019slim} focused on providing frameworks that can facilitate graph analytics with compression scheme. The Ligra framework mainly utilizes the vertex degrees and graph density to select a scheme to map vertices or edges to integer-arrays or bit-vectors,  while SlimGraph focuses more on utilizing statistics of local parts in the input graph to map vertices and edges to higher hierarchies. Thus, its compression kernels can preserve some critical graph properties. However, the use of these frameworks needs to re-program the graph applications with their specific syntax and semantics. This is a non-trivial effort for most of the graph algorithms and applications in the literature.

Moreover, there are a few works that investigate data analytics directly on compressed data without decompression. For example, Zhang {\it et al.} proposed an approach to perform document analytics (word count, inverted index, and sequence count) directly on compressed textual data on CPUs \cite{zhang2021tadoc} and GPUs \cite{zhang2021g, pan2021exploring, liu2022exploring}. 
Moreover, Zhang {\it et al.} developed a new storage engine, called CompressDB, which can support data processing for databases without decompression \cite{zhang2022compressdb}.
Furthermore, Macko {\it et al.} \cite{macko2015llama} proposed LLAMA that performs graph storage and analysis on the compressed sparse row (CSR) representation and achieves performance gain on graph benchmarks (i.e., PageRank, BFS, and triangle counting) due to in-memory execution. 
In addition, Mofrad {\it et al.} \cite{mofrad2019efficient} proposed a compression technique specifically designed for matrix-vector operations based on the compressed sparse column (CSC) representation and leverage this compression to accelerate distributed graph benchmarks (e.g., PageRank, single source shortest path, and BFS). However, no work has been done on using compression to improve memory efficiency and accelerate a real-world diffusion-based graph application such as influence maximization.

\section{Memory Footprint Profiling and Graph Characterization}\label{sec:Profiling}
In this section, we characterize the memory usage of \textcolor{black}{Ripples}, the state-of-the-art parallel IMM implementation 
and discuss the potentials to reduce the memory footprint.

\subsection{Memory Usage of RRR Sets}
The key component of the Monte Carlo (MC) diffusion process  \cite{tang2015influence, minutoli2019fast} is a probabilistic Breadth First Search. The intermediate collections of vertices that may cause activation of the BFS root are saved in the  form of RRR sets. We benchmark six real-world large \textcolor{black}{graphs} used in the literature and study the memory usages for the MC diffusion process and generating/saving the intermediate RRR sets.
As shown in Figure \ref{fig:mem_usage}, the space attributed to storing of the intermediate RRR sets dominate, consuming between 87\% to 99\% of the memory footprint.

\begin{figure}[h]
  \centering
  \includegraphics[width=\linewidth]{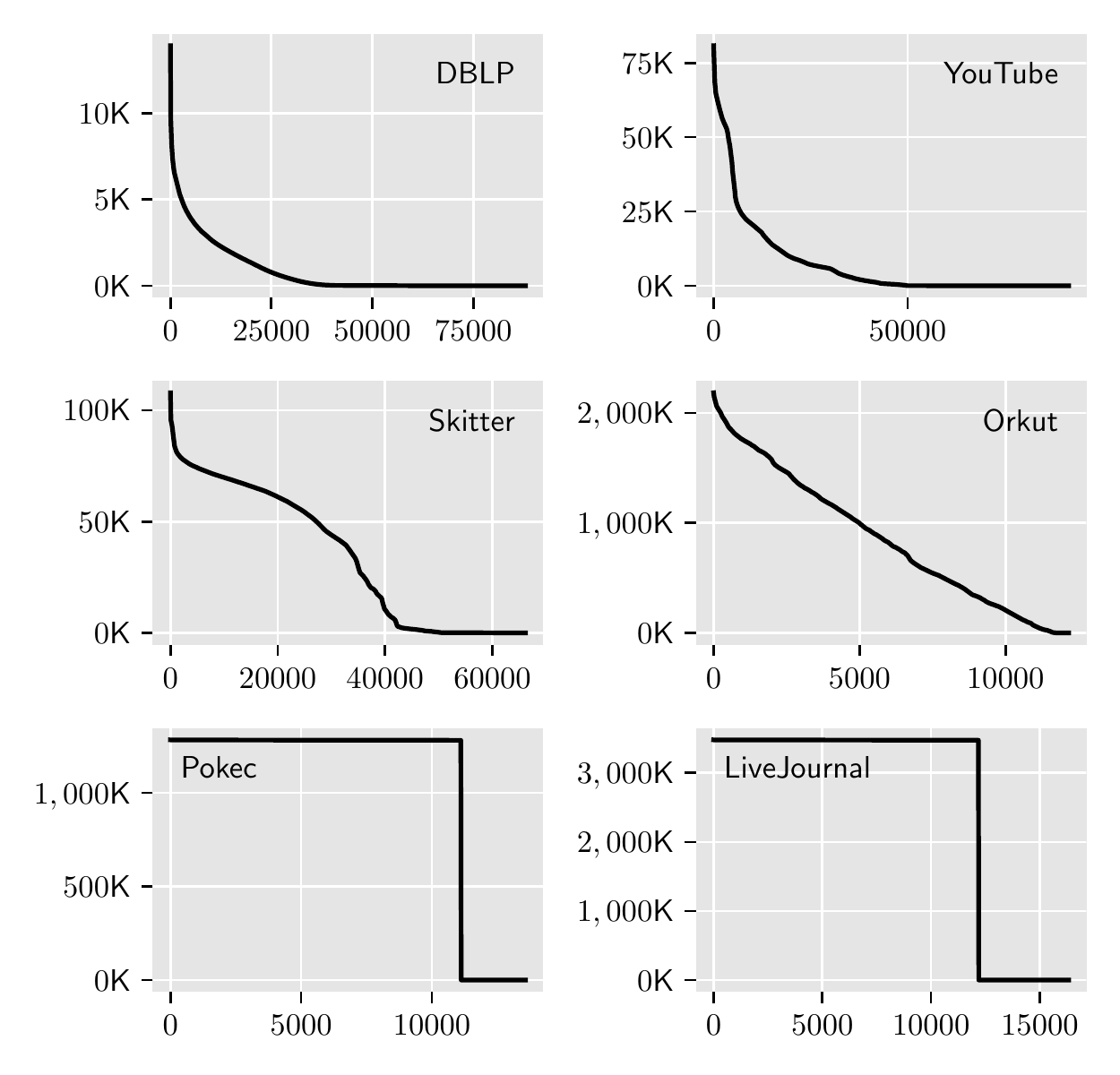}
  \vspace{-4mm}
  \caption{Distributions of RRR set sizes on different graphs (X-axis: the sorted RRR set ID; Y-axis: the number of vertices contained in the RRR set).}
  \vspace{-2mm}
  \label{fig:rrr_hist} 
\end{figure}

To further understand the memory footprint usage patterns in Figure~\ref{fig:mem_usage}, we studied the distribution of the RRR sets by their sizes. 
Figure \ref{fig:rrr_hist} shows the sizes of RRR sets, \add{i.e., their vertex counts,} for different graphs.
The figure illustrates the shapes of RRR sets' distribution based on their sizes:
(1) In the top row, the two graphs (i.e., DBLP and YouTube) have long tails, and their shapes correspond to a power-law distribution. 
(2) In the middle row, the shapes of the two graphs (i.e., Skitters and Orkut) show close to a linear decay without any long tails. 
(3) In the bottom row, the shapes of the two graphs (i.e., Pokec and LiveJournal) show a very uniform two-step (flat-head) distribution.
The overall memory footprint of the RRR sets corresponds to the area under the curves in the respective plots.  
This well explains why the last two graphs spends much more memory in saving the RRR sets.
Further, these observations raise a critical question to the design of optimizations for memory footprint reduction:
\add{What are the characteristics of the various RRR sets of the above three different categories? (see Section~\ref{sec:characteristics}) }

\subsection{Characterize The RRR Sets}
\label{sec:characteristics}
The shapes in Figure~\ref{fig:rrr_hist} roughly categorize the RRR sets' distributions into two types: the first four graphs (i.e., DBLP, YouTube, Skitter and Orkut) have skew-distributed distribution while the last two graphs (i.e., Pokec and LiveJournal) have a uniform two-step distribution. We use two quantities, i.e., skewness and density, to characterize the shapes of RRR sets' distributions. 

\begin{table}
  \caption{Skewness $\mathcal{S}$ and density of RRR sets distributions.}
  \label{tab:skewness}
\begin{tabular}{@{}lcc @{}}
	\toprule
	\TabHead Graph  & Skewness $\mathcal{S}$     & Density $\mathcal{D}$   \\
	\midrule
	DBLP           & \makebox[3em][r]{ $11.46$}$\,\pm\, 0.15$ 
	               & \makebox[3em][r]{ $0.261\%$}$\,\pm\, 0.001\%$ \\ 
	Youtube        & \makebox[3em][r]{  $9.01$}$\,\pm\,  0.06$ 
	               & \makebox[3em][r]{ $0.630\%$}$\,\pm\, 0.001\%$ \\ 
	Skitter        & \makebox[3em][r]{  $5.38$}$\,\pm\,  0.01$ 
	               & \makebox[3em][r]{ $2.030\%$}$\,\pm\, 0.001\%$ \\ 
	Orkut          & \makebox[3em][r]{  $0.75$}$\,\pm\, 0.03$ 
	               & \makebox[3em][r]{ $27.73\%$}$\,\pm\, 0.03\%$ \\ 
	Pokec          & \makebox[3em][r]{ $-1.43$}$\,\pm\, 0.01$ 
	               & \makebox[3em][r]{ $66.01\%$}$\,\pm\, 0.111\%$ \\ 
	LiveJournal    & \makebox[3em][r]{ $-0.99$}$\,\pm\, 0.01$ 
	               & \makebox[3em][r]{ $53.28\%$}$\,\pm\, 0.273\%$ \\ 
	\bottomrule
\end{tabular}
\end{table}

The skewness score ($\mathcal{S}$) measures how distributions are asymmetric about their mean: When $\mathcal{S}<0$, the distribution has a longer left tail; When $\mathcal{S}>0$, the distribution's right tail is longer. Equation (\ref{eq:skewness}) shows the calculation of skewness score for the size of RRR sets $X=(X_1, ...., X_{\theta})$, where \addd{each $X_i$ is a sample of the number of visited vertices starting from a seed vertex in a MC simulation. Thus all $X_i$'s are at least $1$ because there will be no empty resultant RRR set (it will at least contain the starting seed vertex)}; $\theta$ is the total number of sampled RRR sets; $\bar{X}$ is the average RRR size, and $s$ is their standard deviation.\footnote{\addd{The standard deviation $s \neq 0$ because $X_i$'s are the sizes of RRR sets, which are not all equal. Given the elements in Equation (\ref{eq:skewness}) are all non-zero, the skewness score and density will not overflow by dividing zeros for real-world social networks.}} In Table~\ref{tab:skewness}, we show the skewness scores of the first four graphs (i.e., DBLP, YouTube, Skitter and Orkut) are all positive, which help to distinguish them from the last two graphs (i.e., Pokec and LiveJournal) whose skewness scores are negative. 

The density ($\mathcal{D}$) measures the proportion of non-zero elements that are required to represent the sampled RRR sets if they are stored in matrix format. Although density is not correspond to the shape of distributions, it can help to select the data structure to store RRR sets.  It is sufficient to use 32-bit unsigned integers to represent each vertex in the sampled RRR sets in this study. Thus we can use $3.12\%$ to be the threshold. When the density $\mathcal{D} \le \frac{1}{32} = 3.12\%$, it is more efficient to use a sparse representation of RRR sets, i.e., to explicitly store each vertex. On the other hand, when the density $\mathcal{D} > 3.12\%$, it is more efficient to use a dense bitmap coding to store RRR sets. Equation (\ref{eq:skewness}) shows the calculation of density for the size of RRR sets $X$ (Similarly, $\theta$ is the total number of RRR sets; $X_i$ is the number of vertices in each RRR set; $n$ is the number of vertices in graph $G$). In Table~\ref{tab:skewness}, we show the last two graphs (Pokec and LiveJournal) have much higher densities ($\ge 50\%$) than the rest four graphs (DBLP, YouTube, Skitter, Orkut). 
\begin{equation}\label{eq:skewness}
    \mathcal{S} = \frac{\frac{1}{\theta}\sum_{i=1}^{\theta}(X_i-\bar{X})^3}{s^3}, \; \mathcal{D} = \frac{\sum_{i=1}^{\theta}X_i}{\theta \cdot n} 
\end{equation}

\subsection{Characterize The Influence of Vertices}
\label{sec:VertexInfluence}
The different shapes of RRR sets' distribution also characterize the influence of individual vertices which can help us to find data localities. To verify this intuition,  we compare the selected seeds of each graph by varying the random starting vertex for MC samplings. Table \ref{tab:head-seed} shows the Rank-Biased Overlap (RBO) scores \cite{webber2010similarity} ($1.0$ means highly overlapping rank positions and $0.0$ means no overlapping). For the graphs with skew-distributions, the same top-1 influential vertex is consistently selected from different random starting vertices.  However, for the graphs with flat-head distributions, many vertices achieve the maximum influence, so the RBO score is zero.

\begin{table}[t]
  \caption{Influential seeds with a random start.}
  \label{tab:head-seed}
\begin{tabular}{@{} lccc @{}}
	\toprule
	\TabHead Graph
	        &
	\TabHead Activated
	        &
	\TabHead \begin{tabular}{@{}c@{}} RBO \\(Top-1)\end{tabular}
	        &
	\TabHead \begin{tabular}{@{}c@{}} RBO \\(Top-50)\end{tabular} \\
	\midrule
	DBLP    & 0.499 & 1.0 & $0.57\phantom{0}$                          \\
	YouTube & 0.554 & 0.87 & $0.24\phantom{0}$                          \\
	Skitter & 0.794 & 0.50 & $0.16\phantom{0}$                                    \\
	Orkut   & 0.967 & 0.46 & $0.09\phantom{0}$                                    \\
	Pokec   & 0.817 & 0.0 & $0.0\phantom{0}$                                    \\
	LiveJournal & 0.741 & 0.0 & $0.0\phantom{0}$                          \\
	\bottomrule
\end{tabular}
\end{table}

From the first four graphs (DBLP, YouTube, Skitter, Orkut), the sampled RRR sets are close to power-law or linear decay distributions. Under this situation, there are only a small number vertices that influence a lot of other vertices; while many vertices only influence one or two other vertices. 
On the other hand, for the last two graphs (Pokec and LiveJournal), the flat-head shows that many vertices can influence many other vertices. In other words, many vertices are equally influential, which makes the flat-head distributed RRRs lack the data locality in the skew-distributed RRR sets. Thus, they need different optimization strategies to reduce their memory footprints.


In summary, the memory required to store intermediate RRR sets is one to two order of magnitude larger than the memory footprint of the diffusion process.  We observe two types of graphs with regards to their different behaviors in diffusion process and hence \add{the data localities}. 
Considering the observed skewed-distributed RRR sets, it is promising to leverage variable-length encoding to reduce the memory footprint of RRR sets for those graphs. \add{On the ther hand, using bitmap coding can reduce the memory footprints for the flat-head distributed RRR sets because of their high density of non-zero elements.}

\section{Our Proposed Optimizations}\label{sec:Method}
In this section, we propose a \add{``compress-to-compute''} IMM method to reduce the memory footprint for the Ripples based on our characterization and profiling.

\begin{figure}[t]
  \centering
  \includegraphics[width=\linewidth]{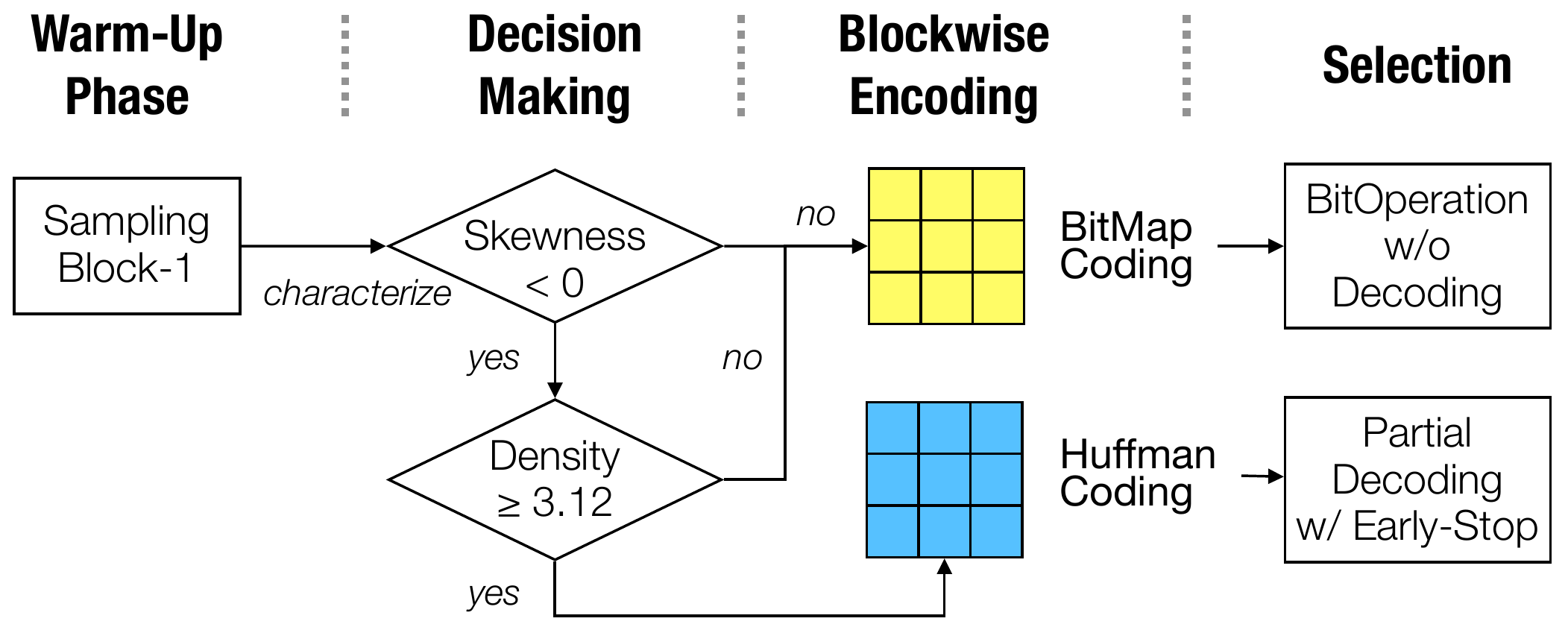}
  \caption{Workflow overview of our proposed solution.}
  \label{fig:flow}
\end{figure}

\subsection{Overview of Our Proposed Workflow}
The overview of the \add{``compress-to-compute''} workflow is shown in Figure \ref{fig:flow}. It contains three main phases: warm-up, sample-and-encode, and decode-and-select.
Specifically, we first characterize the \add{skewness score and the density} from the distribution of the RRR sets in the warm-up phase. Based on that, we then \add{choose an efficient compression technique to encode the intermediate RRR sets and query on the encoded RRR sets.}

More specifically, we introduce a block-based \textit{sampling-and-encoding} approach and an integrated \textit{selection} approach. To achieve a better scalability on multicore architectures, we design two schemes - \add{(1) {\huffmax}: We use Huffman coding to encode skew-distributed RRR sets, and leverage the data locality knowledge acquired from the warm-up phase to accomplish the selection without fully decoding the compressed RRR sets. (2) {\bitmax}: We use bitmap coding to encode flat-head distributed RRR sets. The corresponding selection can be accomplished directly on the encoded data by efficient bit operations.} \addd{These two schemes are complementary to each other so that the proposed workflow does not rely on the data locality of skewed distributions.  It can work with RRR sets that have non-skewed distributed sizes such as normal distributions or uniform distributions (their skewness scores are zero). In Figure~\ref{fig:rrr_hist}, we show the flat-headed distributions of Pokec and Livejournal are two special cases that are close to uniform distributions. } \addd{Furthermore,} we use \textcolor{black}{a heuristic approach} to optimize the parallel reduction operation to find \textcolor{black}{the vertex with the highest frequency}. 
We use Table \ref{tab:notation} to list the notation used in the paper.

\begin{table}[b]
  \caption{Notation used in the paper.}
  \label{tab:notation}
  \resizebox{.95\linewidth}{!}{%
 \begin{tabular}{@{} ll @{} }
    \toprule
    \TabHead Notation & \TabHead Description\\
    \midrule
    $n$ & The number of vertices in Graph \\
    RRR & Random Reverse Reachable set \\
    seeds & The selected most influential vertices \\
    $k$ & The number of most influential vertices in seeds \\
    $\theta$  & The number of random RRR sets to be sampled \\
    $\mathcal{S}$ & \addd{The skewness score of RRR sets}\\
    $\mathcal{D}$ & \addd{The proportion of non-zero elements of sampled RRR sets}\\
    $X_i$ & \addd{The number of vertices in a sampled RRR set}\\
    $\epsilon$ & The approximation factor \\
    $b$ & The number of blocks that consists the Sampling step \\
    $\mathbb{R}_i$ & A block of RRR sets \\
    $rr_j$ & One RR set from the block \\
    $\mathbb{C}_i$ & A block of Huffman encoded RRR sets \\
    $c_j$ & The encoded part of one RRR \\
    $\mathbb{CP}_i$ & A block of copied buffers \\
    $cp_j$ & The copied part of one RRR \\
    $H^*$ & The Huffman codebook \\
    $\hat{h}$ & The shared vertex-frequency table  \\
    $\mathbb{B}_i$ & A block of bitmap encoded RRR sets \\
    $u^*$ & The current most influential vertex \\
  \bottomrule
\end{tabular} 
}
\end{table}


\subsection{Sampling-and-Encoding}
Next, we describe the proposed two encoding methods.

\vspace{-2mm}
\subsubsection{Block-based sampling-and-encoding}
We adopt a block-by-block sampling strategy and use the first block as the warm-up phase. This design provides three benefits: First, as aforementioned, the warm-up phase characterizes the graph and enables us to \add{choose between Huffman coding and bitmap coding.}
Second, the learned characteristics of RRR sets distribution also helps to determine whether the \add{two encoding methods} can efficiently compress RRR sets and hence reduce the memory usage of RRR sets. For RRR sets with negative skewness, using the Huffman coding \add{will cause low compression ratio, and the decoding cannot early stop due to the lack of data localities}; \add{For RRR sets with density $< 3.12\%$, the bitmap coding will even increase the memory footprints.
Third, the block-by-block sampling enables us to interleave the sampling and encoding so that we can release the used memory as soon as possible to maximize the peak memory reduction.}

\vspace{-2mm}
\subsubsection{Encoding of \huffmax}
\label{sec:huffman_encoding}
We simply check the skewness in the warm-up phase to enable the Huffman coding. Table \ref{tab:skewness} shows the skewness scores of our benchmark graphs. The skewed RRR sets of the first four graphs have $\mathcal{S}>0$. They will be encoded with Huffman Codes. Next, we will describe the block-based sampling-and-encoding approach where the Huffman coding method is \add{interleaved with the sampling step}.  Algorithm \ref{alg:block_sampling_compressing} presents the block-based sampling-and-encoding method in detail.

Specifically, we split the sampling step into $b$ sub-steps. In each sub-step we get a block of $\frac{\theta}{b}$ RRR sets. Let $\mathbb{R}_1 = \{rr_1,...,rr_{\theta/b}\}$ be the first block. 
We build a Huffman codebook $H^*$ based on the block $\mathbb{R}_1$. 
We use this codebook $H^*$ to encode $\mathbb{R}_1$ into a set of byte-strings $\mathbb{C}_1=\{c_1,...,c_{\theta/b}\}$, each of which corresponds to a RRR set. 
At the same time, we construct a frequency table $\hat{h}$ to store the vertex frequency of the block $\mathbb{R}_1$. 
In the following sub-steps, we keep updating the frequencies stored in the table $\hat{h}$ and encoding the RRR sets $\mathbb{R}_i$ into $\mathbb{C}_i$, where $i=2, \cdots, b$. Because \add{of the data localities of the skew-distributed RRR sets}, we can keep track of the current most influential vertex $u^*$ \add{and swap it to} the beginning \add{position} of the encoded byte string \add{whenever} it appears in the corresponding RRR set. \add{Note that this swap is beneficial as it enables early-stopping (will be described in Section \ref{sec:selection}).} At the end of each sub-step, \add{to reduce the peak memory usage}, the memory for $\mathbb{R}_i$ is deallocated once the encoding of the current block completes.

Ideally, the codebook $H^*$ built from the first block $\mathbb{R}_1$ should contain the Huffman codes for all vertices in the entire RRR sets. 
However, in our block-based sampling-and-encoding approach, some vertices may not be sampled in $\mathbb{R}_1$ but could appear in later blocks \addd{due to the skewed distributions. The more skewed distributions the RRR sets have, the more likely some vertices are missing from $\mathbb{R}_1$. For example, 
0.2\% of DBLP's vertices have no Huffman-codes where the skewness score is 11.46; 
0.02\% vertices of Skitters have no Huffman-code where the skewness is 5.38; 
and 0.003\% vertices of Orkut’s have no Huffman-code where the skewness is 0.75.
} 
For those vertices do not have corresponding Huffman codes in the codebook $H^*$, we use an additional array $\mathbb{CP}_i$ to directly save those vertices without encoding.
By doing this, we \add{preserve the exact information of RRR sets by} encoding and/or copying all the vertices in the original RRR sets. \add{From our empirical studies, the vertices that do not have corresponding Huffman codes only consist of less than $0.2\%$ of the entire vertices. Thus we do not build new Huffman code book in the following steps to reduce the overhead. }

\begin{algorithm}[t]
\caption{Sampling-and-Encoding($G, \theta, b$)}\label{alg:block_sampling_compressing}
\begin{algorithmic}[1]
\State $\mathbb{B}=\phi$,\quad $\mathbb{C}=\phi$,\quad $\mathbb{CP}=\phi$,\quad $\hat{h}$ = zeros
\For {$i \le b$}
\State $\mathbb{R}_i=\text{Sampling} \big(G, \frac{\theta}{b}, \text{threshold} \big)$
\If{$i \equiv 1$}
    \State $\mathcal{S,D}=$ Characterize($\mathbb{R}_i$)
    \If{$\mathcal{S}<0$}
        \State \bitmax$=True$
    \Else
        \State \huffmax$=True$
        \State $H^* = $ BuildHuffmanCode($\mathbb{R}_i$);
    \EndIf
\EndIf
\If{ \huffmax $\equiv True$}
    \State $\hat{h} = \text{UpdateHistogram} \big(\mathbb{R}_i, \hat{h} \big)$
    \State $[\mathbb{C}_i, \mathbb{CP}_i] = \text{HuffmanEncode}\big(H^*, \hat{h}, \mathbb{R}_i\big)$
    \State  $\mathbb{C}=\mathbb{C} \cup \mathbb{C}_i$;\quad$\mathbb{CP}=\mathbb{CP} \cup \mathbb{CP}_i$
\EndIf
\If{ \bitmax $\equiv True$}
    \State $\mathbb{B}_i = \text{BitmapEncode}\big(\mathbb{R}_i\big)$
    \State  $\mathbb{B}=\mathbb{B} \cup \mathbb{B}_i$
\EndIf
\State Deallocate  $\mathbb{R}_i$
\EndFor
\end{algorithmic}
\end{algorithm}

\vspace{-2mm}
\subsubsection{Encoding of \bitmax}
\add{We not only check the skewness ($<0$), but also check the density ($>3.12\%$) to enable the bitmap coding (as mentioned in Section \ref{sec:characteristics}). For a block of RRR sets $\mathbb{R}_i$, the encoded data is represented as a dense bit matrix $\mathbb{B}_i$, where the shape of matrix $\mathbb{B}_i$ is $n$ rows by $\frac{\theta}{b}$ columns ($n$ is the number of vertices in graph G, $\frac{\theta}{b} $ is the number of RRR sets sampled in this block). If the bit at the $r$-th row and $c$-th column in matrix  $\mathbb{B}_i$ is set to $1$, it represents the $r$-th vertex appears in the $c$-th RRR set. In our implementation, we pad $\frac{\theta}{b}$ columns to be the multiple of $32$ so that we can save the bit matrix in byte format. Because the padded bits are all zeros, the padded columns do not affect the correctness of the method. Similar to \huffmax, 
we deallocate the $\mathbb{R}_i$ after the bitmap encoding to reduce the peak memory usage. }

\vspace{-2mm}
\subsubsection{Parallel implementation}\label{sec:parallel-encoding}
The sampling-and-encoding computation on one sample is independent of other RRR sets. To accelerate the computation, we distribute the sampling-and-encoding workload to multiple threads/cores. For \huffmax, the bottleneck is to perform a parallel reduction to build a shared frequency table $\hat{h}$ considering the NUMA effect  \cite{lameter2013overview}. Our solution is to follow the first-touch principle: we allocate a local frequency table $h^{local}$ on each thread to store the vertex frequencies and sum them to the global frequency table after the sampling-and-encoding of this block completes (the synchronization happens $b$ times as we split the sampling-and-encoding into $b$ sub-steps). \add{For \bitmax, its encoding is embarrassingly parallel on multi-cores for high scalability.}

\subsection{Optimized Selection}  
\label{sec:selection}
Lastly, we describe two selection approaches 
for efficiently querying in \huffmax~ and \bitmax, respectively. 
\add{The selection iterates two core computations for $k$ times to select the most influential seeds. The first computation is to locate and remove the RRR sets which contain the current most frequent vertex $u^*$. The second computation is to reconstruct the frequency table $\hat{h}$ after every RRR set that contains the previous $u^*$ is removed. After that, a new $u^*$ is selected based on the updated frequency table $\hat{h}$.}

\vspace{-2mm}
\subsubsection{Selection of \huffmax}
\add{For RRR sets encoded by Huffman coding, we leverage the data locality to swap the most frequent vertex $u^*$ to the beginning position during the {\huffmax} encoding steps (described in Section \ref{sec:huffman_encoding}). This enables us to partially decode an encoded RRR set (i.e., $c_j$ where $j=1, \cdots, \theta$) and stop early whenever it contains $u^*$.  Due to the skew-distributed RRR sets, we only need to decode a small number of encoded RRR sets. } \addd{As aforementioned in Table~\ref{tab:skewness}, graphs of greater skewness scores (DBLP/11.46, YouTube/9.01) also have longer tails than graphs of smaller skewness scores(Skitter/5.38, Orkut/0.75). However, the difference of tails does not affect the selection efficiency: even graph with small positive skewness (Orkut/0.75) still has the data locality that enables {\huffmax} to early-stop in the selection step.}
\add{Note that when we decode $c_j$ back to a temporary buffer $tmp=\{v_1, \cdots\}$, it implies that the current $u^*$ does not appears in the encoded part of this RRR set; otherwise, we would have early stopped decoding. In this case, we will search the corresponding copied array $cp_j$.}

\begin{algorithm}[t]
\caption{{\huffmax}DecodeQuery$ \big( H^*,\hat{h},\mathbb{C},\mathbb{CP} \big)$}\label{alg:huffmax_select}
\begin{algorithmic}[1]
\State {$u^*$=argmax($\hat{h}$),\quad seeds=$\phi$ ,\quad deleteRRRflag[$\theta$]=$False$}
\While{$|\text{seeds}| \le  k$ }
\State $\hat{h}=$ zeros;\quad {$\text{seeds} = \text{seeds} \cup u^*$}
\While{$j \le  \theta$ }
\If{deleteRRRflag$_j \equiv False$}
    \State tmp, findflag = DecodeFind($H^*,c_j,cp_j,u^*$)
    \If{findflag $\equiv True$}
        \State{deleteRRRflag$_j=True$}    
    \Else
        \State $\hat{h}=\text{UpdateHistogram} \big(tmp,cp_j,\hat{h}\big)$
    \EndIf
\EndIf
\State Deallocate $tmp$
\EndWhile
\State {$u^*$=argmax($\hat{h}$)}
\EndWhile
\end{algorithmic}
\vspace{-6mm}
\end{algorithm}

To reconstruct the new frequency table $\hat{h}$, we only need to count the vertices in the fully decoded buffer $tmp$ and the corresponding copied array $cp_j$ if $u^*$ is not found. At this point, the temporary buffer $tmp$ is safely deallocated to reduce memory footprints. Algorithm \ref{alg:huffmax_select} describes this {\huffmax} selection in detail.

\vspace{-2mm}
\subsubsection{Selection of \bitmax}
\add{
Unlike \huffmax's selection, where we leverage the data localities to partially decode the compressed data, the selection in {\bitmax} directly operate on the encoded data without decoding. The selection is accomplished by efficient bit operations. Note that the shape of bitmap encoded $\mathbb{B}$ is different from the Huffman encoded $\mathbb{C}$; we will use an example to illustrate the two core selection computations (i.e., locate and remove RRR sets which contains $u^*$, reconstruct frequency table to find the new $u^*$).
}
\begin{equation}\label{eq:bitmax-op}
    \begin{pmatrix}
      v_1^*|  1 & 1 & 0 & 0 & 1 & 1\\
      v_2  |  0 & 1 & 1 & 0 & 0 & 1\\
      v_3  |  1 & 0 & 1 & 0 & 1 & 0\\
      v_4  |  0 & 1 & 0 & 1 & 0 & 1\\
      v_5  |  0 & 0 & 1 & 1 & 1 & 0  
    \end{pmatrix}
    \rightarrow
    \begin{pmatrix}
      v_1^*|  0 & 0 & 0 & 0 & 0 & 0\\
      v_2  |  0 & 0 & 1 & 0 & 0 & 0\\
      v_3  |  0 & 0 & 1 & 0 & 0 & 0\\
      v_4  |  0 & 0 & 0 & 1 & 0 & 0\\
      v_5  |  0 & 0 & 1 & 1 & 0 & 0
    \end{pmatrix}
  \end{equation}
  
\add{Equation (\ref{eq:bitmax-op}) shows a simple example that has $5$ vertices and $6$ RRR sets. The frequency table $\hat{h}$ is computed by row-wise \texttt{POPCOUNT} operation (i.e., find the sum of ones in each row). Given their frequencies, $v_1$ is selected as the current most frequent $u^*$. We do not need to locate the RRR sets because their locations are explicitly represented by the $1$s in $v_1$'s row (for simplicity, we use the notation $v_i$ to represent the $v_i$'s row). To remove these RRR sets, we \texttt{subtract} row $v_1$ from all the other rows. }
\add{We use two bit operations to accomplish the \texttt{SUBTRACT} operation on the bitmap-encoded data. Specifically, we use $\texttt{tmp} = v_i \texttt{ AND } (v_i \texttt{ XOR } u^*)$ to \texttt{subtract} row $u^*$ from row $v_i$.}
  
\begin{algorithm}[t]
\caption{{\bitmax}Query$ \big( \hat{h},\mathbb{B}\big)$}\label{alg:bitmax_select}
\begin{algorithmic}[1]
\State {$u^*$=argmax($\hat{h}$),\quad seeds=$\phi$ ,\quad deleteVTXflag[$n$]=$False$}
\While{$|\text{seeds}| \le  k$ }
\State $\hat{h}=$ zeros;\quad {$\text{seeds} = \text{seeds} \cup u^*$}
\While{$i \le  n$ }
\If{deleteVTXflag$_i \equiv False$}
    \State \texttt{SUBTRACT} ($v_i,u^*$)
    \If{\texttt{POPCOUNT}($v_i$) $\equiv 0$}
        \State{deleteVTXflag$_i=True$}    
    \Else
        \State $\hat{h}=\text{UpdateHistogram} \big(\texttt{POPCOUNT}(v_i))$
    \EndIf
\EndIf
\EndWhile
\State {$u^*$=argmax($\hat{h}$)}
\EndWhile
\end{algorithmic}
\vspace{-6mm}
\end{algorithm}

\add{After \texttt{SUBTRACT}, we use row-wise \texttt{POPCOUNT} again on the updated bitmap matrix $\mathbb{B}$ to reconstruct the updated frequency table $\hat{h}$.} Algorithm-\ref{alg:bitmax_select} describes the selection of {\bitmax} in detail.

\subsubsection{Parallel implementation} 
\addd{The above example in Equation (\ref{eq:bitmax-op}) demonstrates the bit operations of \texttt{POPCOUNT}, \texttt{AND}, and \texttt{XOR} on each row of vertices. However, the bit-columns of the encoded bitmap $\mathbb{B}$ are distributed across different threads according to the first touch principle (as described in Section \ref{sec:parallel-encoding}). It is important to consider the NUMA effect to parallelize the selection step of {\bitmax} along the bit-column direction instead of iteratively applying bit operations on each row. More concretely, each thread keeps track of a local frequency table $h^{local}$ and these local frequency tables are reduced to the global frequency table $\hat{h}$ after each iteration. The next section describes our solution to address the scalability challenge on getting the global most frequent vertex $u^*$ without reducing the entire global frequency table $\hat{h}$.}

\subsubsection{Parallel merge} 
\begin{figure}[t]
  \centering
  \includegraphics[width=0.95\linewidth]{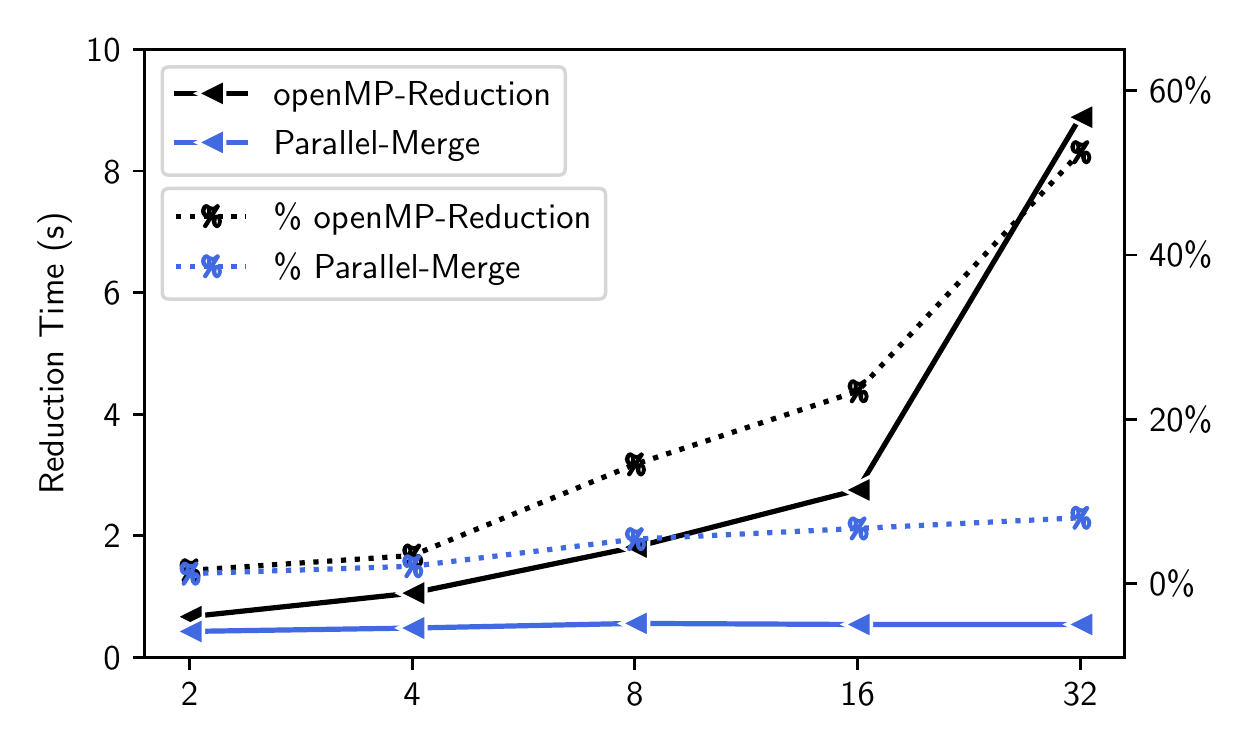}
  \caption{Time comparison of reduction on Skitter's frequency table $\hat{h}$ with different \# of threads (solid line: absolute time; dashed line: relative to the time-to-solution).}
  \label{fig:parallel_merge} 
\end{figure}

\add{Both selections in {\huffmax} and {\bitmax} need to find the new most influential vertex $u^*$ after reconstructing the frequency table.} However, we note that to generate the overall reconstructed frequency table $\hat{h}$ from multiple threads, a parallel reduction with $k$ times of synchronizations is needed \addd{to sum local frequencies for all vertices.} \addd{This reduction} introduces high time overhead, as shown in Figure \ref{fig:parallel_merge}. The reduction times are measured on the Skitter graph. The dashed line shows the runtime of the OpenMP reduction function with 2$\sim$32 threads/cores. It illustrates that the original OpenMP reduction would become the performance bottleneck as the number of threads increases. The reduction time takes upto $52.5\%$ of the entire time-to-solution.

To solve this issue, we propose to use the following approach to \addd{greatly decrease the number of reductions so as to} make the \addd{selection step} more efficient and scalable: \textcolor{black}{(1) Select the locally most frequent vertex from each thread \addd{(these are local maxima)}; 
(2) Perform reductions to get the global frequencies for these locally selected vertices \addd{(the local maxima)};
and (3) Select the vertex with the maximum global frequency from the locally selected vertices \addd{( get the global maximum from these local maxima)}.
Note that the total number of reduction operations in (2) is $k \times p$ instead of $ k \times n$, where $p$ is the number of threads and $n$ is the number of vertices in graph ($p<<n$).
Figure \ref{fig:parallel_merge} illustrates that our proposed parallel-merge-based reduction is highly efficient and scalable with almost constant time cost.} \textcolor{black}{In this example, the number of vertices of Skitter is about 1.6 millions. 
With 32 threads, the OpenMP reduction needs to reduce 650 MB data, whereas our approach only needs to reduce 12.5 KB data\footnote{In the skitter example (n=1.6M, k=100), each frequency needs 4 bytes (one float), the baseline reduction needs to sum 1.6M*100*4=650MB data; and with 32 threads, our proposed method sums 32$\times$100$\times$4=12.5 KB data.}. This explains why our approach is more efficient with the constant computation time of about 0.5 seconds.}

\textcolor{black}{The heuristic is based on the fact that (1) The global RRR sets is the collection of local RRR sets from each thread. The local RRR sets on each thread is independent because the MC simulations are uniformly sampled. (2) Let $P_v(X)$ be the distribution of vertex $v$'s frequency in the global RRR sets. 
Then, the distribution of vertex $v$'s frequency on each thread is $P_v(X/p)$. (3) Given two vertices $u,v$, the corresponding distributions of their frequencies are $P_u(X), P_v(X)$ if they are measured from the global RRR sets. Let $P_u(X) \ge P_v(X)$, without loss of generality, we also have $P_u(X/p) \ge P_v(X/p)$ on each thread. (4) Thus the globally most frequent vertex is among the locally most frequent vertices from each thread.}

\section{Experimental Evaluation}\label{sec:Evaluation}
In this section, we first present our experimental setup. We then demonstrate the effectiveness of our \add{``compress-to-compute''} method. After that, we show the performance of {\hbmax} in memory usage and computation time. Finally, we show that our method has strong scalability on the multicore architectures.

\subsection{Experimental Setup}
\paragraph{Test datasets} For evaluation purpose, we use the five largest graphs tested in the Ripples software \add{and three large graphs not tested by Ripples\footnote{Tested graphs are from the the SNAP collection \cite{snapnets}, the UFL sparse matrix collection \cite{davis2011university}, and Laboratory for Web Algorithms (LAW) \cite{BRSLLP,BoVWFI}.}}. Table \ref{tab:data} shows the main features of these widely studied networks. In detail, (1) DBLP \cite{yang2015defining} describes a co-authorship network of researchers in computer science; (2) YouTube \cite{mislove2007measurement} is a social network of friendship and groups on the video-sharing web site; (3) Skitter \cite{caida:skitter} is an internet topology graph by depicting the forward paths that actual packets traversed to a destination; (4) Orkut \cite{mislove2007measurement} is generated from an on-line social networking site where its primary purpose is to finding and connecting new users; (5) \textcolor{black}{Pokec \cite{takac2012data} is an on-line social network in Slovakia and Czech Republic}; (6) LiveJournal \cite{backstrom2006group} is a graph from a social networking and blogging site, with explicit user-defined communities; \add{(7) Arabic-2005 \cite{BoVWFI,BRSLLP,BCSU3} crawls web pages written in Arabic; and (8) Twitter7 \cite{yang2011patterns} contains the graph of Twitter posts covering a 7-month period.} \add{The first six graphs are characterized (as shown in Figure \ref{fig:rrr_hist}) to help us design {\hbmax}, while the last two are not characterized beforehand. We use these two graphs as a test set to verify the effectiveness of {\hbmax}. For the Arabic graph, $\mathcal{S}=-0.25$, $\mathcal{D}=0.22$, and for the Twitter7 graph, $\mathcal{S}=-3.19$, $\mathcal{D}=0.62$. Thus, these two graphs use {\bitmax}.} \addd{Note that the input data of the two largest graphs uses large amount of storage. Specifically, Arabic has over 22.7 million vertices and 639.9 million edges (requiring more than 11 GB storage overhead), while Twitter7 has 41.6 million vertices and 1.46 billion edges (requiring more than 23 GB storage overhead). These two largest graphs cannot be run by Ripples on our tested platforms due to memory overflow.}

\begin{table}[t!]
  \caption{Input real-world graphs tested.}
  \label{tab:data}
  \resizebox{0.86\linewidth}{!}{%
  \begin{tabular}{@{} lrrrr @{}}
	\toprule
	\bfseries Network & \bfseries \#Vertices & \bfseries \#Edges & \bfseries Avg Deg & \bfseries Max Deg \\
	\midrule
	DBLP              & 317,080              & 1,049,866         & 3.31                 & 306                  \\
	YouTube           & 1,134,890            & 2,987,624         & 2.63                 & 28,576               \\
	Skitter           & 1,696,415            & 11,095,298        & 6.54                 & 35,387               \\
	Orkut             & 3,072,441            & 117,185,083       & 76.28                & 33,313               \\
	Pokec             & 1,632,803            & 30,622,564        & 37.51                & 20,518               \\
 	LiveJournal       & 4,847,571            & 68,993,773        & 28.47                & 22,889              \\
 	arabic-2005       & 22,744,080           & 639,999,458       & 28.14                & 575,618             \\
 	twitter7          & 41,652,230           & 1,468,365,182     & 35.25                & 770,155             \\
	\bottomrule
\end{tabular}
  }
  \vspace{-4mm}
\end{table}

\vspace{-2mm}
\paragraph{Evaluation platforms} We evaluate our proposed method and compare it with the baseline \addd{on two platforms. The first platform is a local workstation which has 4 Intel Xeon Gold 6238R CPUs, 376 GB RAM and 1 TB NVMe SSD storage. This local workstation is used to make a fair comparison with Ripples because it has larger memory.} \addd{The second platform is a} Regular Memory (RM) compute node from the Bridges-2 supercomputer \cite{bridge2} at PSC, which has 2 AMD EPYC 7742 CPUs (64 cores per CPU) and 256 GB RAM. \addd{The Bridges-2 compute node is used for scalability study because it has more CPU cores.} \textcolor{black}{We use GCC-8.3.1 for compilation.} 

\vspace{-2mm}
\paragraph{Parameter setup} We choose $k=100$ for all the graphs. We use $\epsilon=0.2$ for DBLP, YouTube, and Skitter, $\epsilon=0.5$ for Orkut, Pokec, and LiveJournal, and $\epsilon=0.7$ for Arabic and Twitter based on the sizes of these graphs. Note that it takes 24+ hours to run Twitter7 using 2 cores on Bridges2.

\vspace{-2mm}
\paragraph{Our implementation and comparison baseline} 
We use \hbmax~to denote our solution; specifically, we use \huffmax~to denote our solution using the \add{Huffman encoding with partial decoding for selection} and \bitmax~to denote our solution using the \add{bitmap encoding with non-decoding selection}. 
\textcolor{black}{We compare \add{\hbmax} with the original \texttt{Ripples} v.2.1.
We implement our optimization in parallel based on this version of Ripples by using OpenMP-4.1.1.
Note that \texttt{Ripples} provides a flexibility to configure the number of threads for sampling and selection phases separately. In our evaluation, {\hbmax} uses the same number of threads for both phases, \addd{while \texttt{Ripples} can use different number of threads in the selection phase to get the best performance.} 
}

\begin{table}
  \caption{\addd{Sampling time (in seconds) and reduced page faults (in percentile) of {\ripples} and {\hbmax}.}}
  \label{tab:samptime}
  \resizebox{0.75\linewidth}{!}{%
  \begin{tabular}{@{} lrrc @{}}
	\toprule
	\bfseries Graph & \bfseries \hbmax & \bfseries Ripples & \bfseries Reduced Page Faults \\
	\midrule
	DBLP              & 0.45           & 0.50    &   $29.5\%$  \\
	YouTube           & 4.30           & 5.78    &   $24.3\%$  \\
	Skitter           & 13.43          & 17.56   &   $10.5\%$  \\
	Orkut             & 196.02         & 245.25  &   $25.2\%$  \\
	Pokec             & 196.79         & 257.96  &   $31.0\%$  \\
 	LiveJournal       & 612.60         & 761.59  &   $42.8\%$  \\
 	arabic-2005       & 1297.60        & NA      &  NA  \\
 	twitter7          & 11219.10       & NA      &  NA  \\
	\bottomrule
\end{tabular}
  }
\end{table}

\begin{table*}
  \caption{Memory footprint (in MB) and reduction ratio  (shown in \add{parenthesis}).}
  \label{tab:memory}
  \resizebox{0.7\linewidth}{!}{%
  \begin{tabular}{@{}lcccccccc@{}}
	\toprule
	\TabHead Graph & \TabHead DBLP & \TabHead YouTube & \TabHead Skitter & \TabHead Orkut & \TabHead Pokec & \TabHead Journal & \TabHead Arabic & \TabHead Twitter7\\
	\midrule
	Ripples        & 424 (1.00)    & 3,143 (1.00)     & 9,838 (1.00)     & 46,506 (1.00) & 55,682 (1.00) & 163,745 (1.00)  & \cellcolor{gray!30} 348,606(1.00) & \cellcolor{gray!30} 1,193,006(1.00) \\
	\huffmax          & 316 (1.34)    & 1,722 (1.83)     & 5,293 (1.86)     & 30,130 (1.54)   & -   & - & - & - \\
	\bitmax        & -    & -     & -     & - & 10,661 (5.22)   & 29,329 (5.58) & 81,504(4.28)  & 200,250(5.96) \\
	\bottomrule
\end{tabular}
  }
\end{table*}

\begin{table*}
  \caption{Time-to-solution (in second) and overhead ratio  (shown in \add{parenthesis}).}
  \label{tab:time}
  \resizebox{0.7\linewidth}{!}{%
  \begin{tabular}{@{}lcccccccc@{}}
	\toprule
	\TabHead Graph & \TabHead DBLP & \TabHead YouTube & \TabHead Skitter & \TabHead Orkut & \TabHead Pokec & \TabHead Journal & \TabHead Arabic & \TabHead Twitter7\\
	\midrule
	Ripples        & 0.95 (1.0)    & 6.95 (1.0)     & 20.46 (1.0)     & 249.35 (1.0) & 262.66 (1.0) & 775.58 (1.0)  & NA & NA \\
	\huffmax          & 1.10 (1.16)    & 6.31 (0.91)     & 17.93 (0.88)     & 235.14 (0.94)   & -   & - & - & - \\
	\bitmax        & -    & -     & -     & - & 222.63 (0.85)   & 692.70 (0.89) & 1,608.48  & 12,098.30 \\
	\bottomrule
\end{tabular}
  }
\end{table*}

\begin{table*}
  \caption{Time-to-solution (in second) and overhead ratio  with the same memory footprint.}
  \label{tab:io}
  \resizebox{0.75\linewidth}{!}{%
  \begin{tabular}{@{}lcccccccc@{}}
	\toprule
	\TabHead Graph & \TabHead DBLP & \TabHead YouTube & \TabHead Skitter & \TabHead Orkut & \TabHead Pokec & \TabHead Journal & \TabHead Arabic & \TabHead Twitter7\\
	\midrule
	Ripples        & 1.68 (1.0)    & 20.69 (1.0)     & 85.95 (1.0)     & 669.85 (1.0) & 998.85 (1.0) & 1930.60 (1.0)  & 5463.62(1.0) & 26825.60(1.0) \\
	\hbmax          & 1.10 (0.66)    & 6.31 (0.30)     & 17.93 (0.21)     & 235.14 (0.35)   & 222.63 (0.22)   & 692.70 (0.36) & 1608.48 (0.29) & 12098.30 (0.45) \\
\bottomrule
\end{tabular}
  }
\end{table*}

\begin{figure*}
\includegraphics[width=0.85\linewidth]{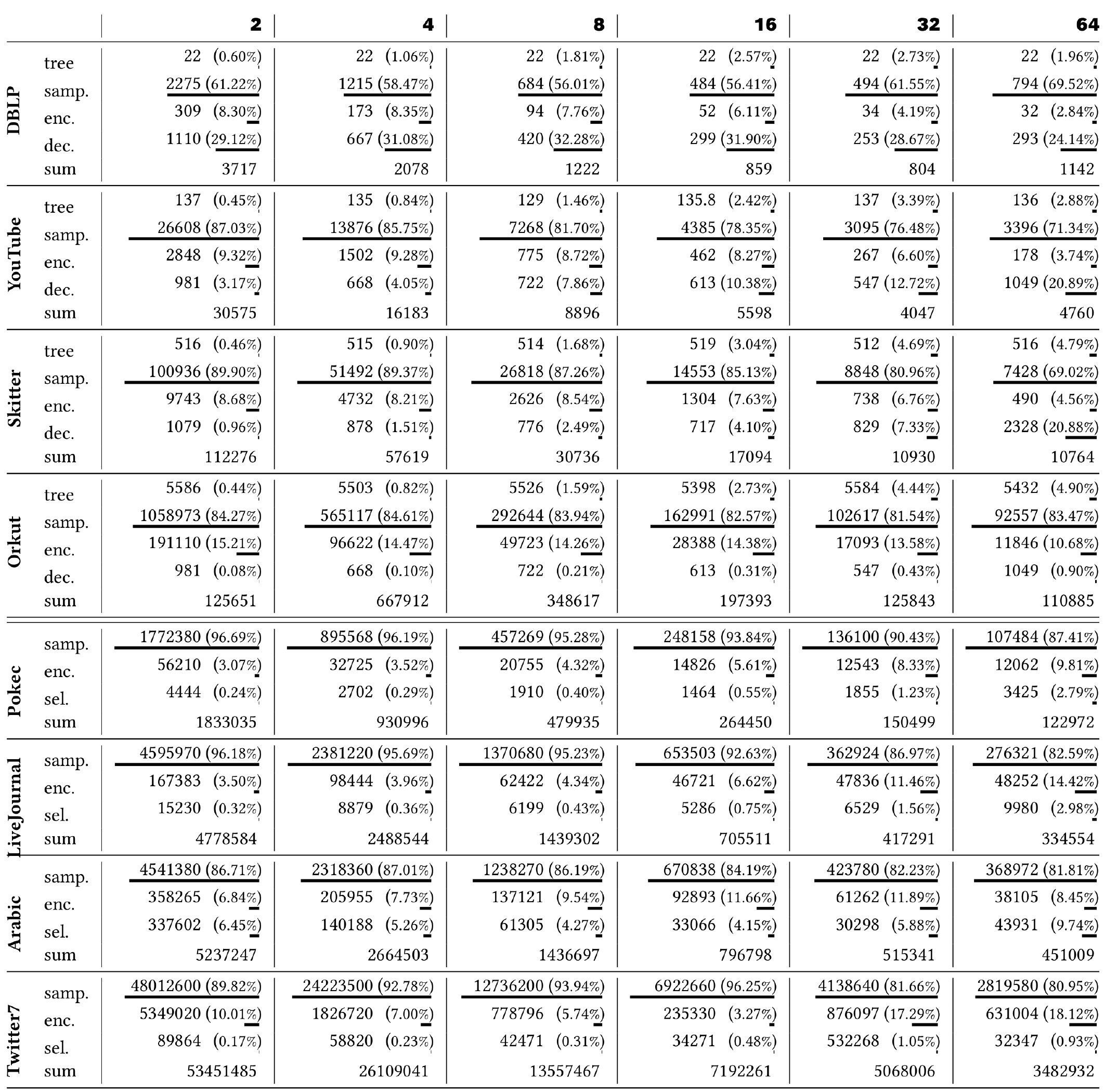}
\caption{Time breakdown of \hbmax~with different numbers of threads/cores on our tested graphs. \add{The underscore's length of each operation is proportional to its runtime.}
}
\vspace{-4mm}
  \label{fig:breakdown}
\end{figure*}

\subsection{Performance Evaluation}
\begin{figure*}[h]
  \centering
    \includegraphics[width=\linewidth]{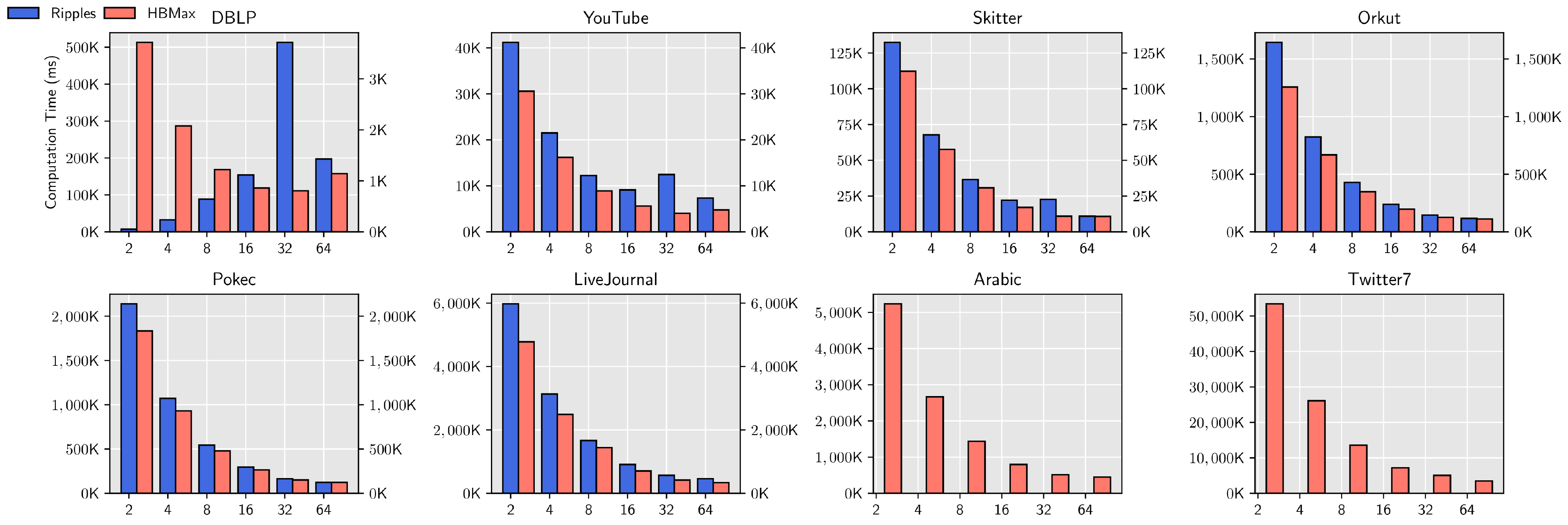}    
  \caption{Both \hbmax~ and Ripples shows strong scalability on tested graphs, except Ripples does not scale on the DBLP and YouTube graph. Note that Ripples does not run on the last two graphs - Arabic and Twitter7 because of OOM error. }
  \vspace{-2mm}
  \label{fig:scale1}
\end{figure*}

In this section, we evaluate the performance of our proposed {\hbmax}. 
\addd{We first evaluate on the local workstation to compare the memory usage and time-to-solution of {\hbmax} with \texttt{Ripples} because it has larger memory (376 GBs). The experiment results are measured with 16 cores.  For a fair comparison, we repeat the experiments on each tested graph for five times and take the averages to compare memory usage and time-to-solution of {\hbmax} with \texttt{Ripples}. We show the averaged experiment results in Table~\ref{tab:memory}, \ref{tab:time} and \ref{tab:io}. We then use Bridges-2 to study the scalability of {\hbmax} because its compute node has more CPU cores (i.e., 64 cores per CPU). We show the scalability results in Figure~\ref{fig:breakdown} and \ref{fig:scale1}. } 
\vspace{-2mm}
\subsubsection{Memory Reduction Evaluation}
Table \ref{tab:memory}  shows the memory footprint and memory reduction ratio of our proposed solution compared with \texttt{Ripples}. The reduction ratio is computed as the ratio of our memory usage to \texttt{Ripples}'s memory usage.
With the proposed \add{``compress-to-compute''} techniques, {\huffmax} achieves an average of 1.6$\times$ memory reduction \add{on four skew-distributed RRR sets (i.e., reduces $39.1\%$ memory footprint); {\bitmax} achieves an average of 5.3$\times$ memory reduction (i.e., reduces $81.0\%$ memory footprint)\add{on two flat-head distributed RRR sets}. Note that the memory usage of \texttt{Ripples} are colored to grey for the last two graphs (Arabic and Twitter7). The numbers are projected from accumulation. The original \texttt{Ripples} cannot finish due to out-of-memory (OOM) error}.

\vspace{-2mm}
\subsubsection{Time-to-Solution Evaluation}
\addd{Table~\ref{tab:samptime} shows {\hbmax} has better performance than {\ripples} in the sampling phase. This is because, although the sampling phases of {\hbmax} and \texttt{Ripples} are identical ({\hbmax} only optimizes the storage of intermediate RRR sets in the seed selection phase), reducing memory footprint can decrease the page faults for the sampling phase (the average reduced percentile of page faults is $27.2\%$\footnote{\addd{the reduced page fault percentile $= (\text{PF}_{\ripples}-\text{PF}_{\hbmax})/\text{PF}_{\ripples} \times 100\%$}}) and hence shorten the sampling time and the overall time-to-solution. In other words, the encoding/decoding overheads are offset by the performance gain from the improved sampling phase.} 

Table \ref{tab:time} shows the time-to-solution of our proposed solution compared with \texttt{Ripples}. We calculate the overhead (in ratio) as $Time_{ours}/Time_{base}$. When our time-to-solution is longer than that of \texttt{Ripples}, the overhead is greater than $1.0$; otherwise, it is less than $1.0$. 
The table shows that \huffmax~ achieves an average overhead of $0.97$ for the \add{four skew-distributed RRR sets}, while the \bitmax~ achieves an average overhead of \add{$0.87$ for the two flat-head distributed RRR sets}.
In other words, {\huffmax} reduces the time-to-solution \addd{and the memory footprints at the same time.}
\add{More concretely, the average reduction on peak memory usage is $52.5\%$ with $6.3\%$ \addd{speedup} on time-to-solution on all the tested graphs (not including the last two graphs that \texttt{Ripples} gets OOM error). Our solution reduces the peak memory usage by up to $82.1\%$ (on LiveJournal) and the time-to-solution by up to $15.0\%$ (on Pokec).}

Moreover, the memory reduction also brings performance benefits to resource-constrained systems. 
To simulate such an environment, we simply limit the amount of available memory for \texttt{Ripples} to be the same as the memory usage by \hbmax. We modify \texttt{Ripples} to let it write the exceeded RRR sets during the sampling step to external SSD storage and read them back for the seed selection step. 
Table \ref{tab:io} shows the performance gain of \hbmax~ over \texttt{Ripples} with the same limited memory capacity. The in-memory compression technique increases the overall performance on the eight skewed-distributed graphs. The average performance gain is $64.4\%$, which equals 3.17$\times$ speed-up. This is obviously because \texttt{Ripples} increases I/O time due to insufficient memory, while our solution saves RRR sets in a compressed format to avoid data movement.

\subsection{Strong Scalability Evaluation}
We now present the strong scaling result of {\hbmax} with up to 64 CPU cores, as shown in Figure \ref{fig:scale1}.
We break down {\hbmax} into four operations:  (1) sampling the RRR sets with MC diffusion process, (2) encoding the RRR sets with {\huffmax}, (3) encoding the RRR sets with {\bitmax} method, and (4) selecting the most influential seeds. We analyze their impacts to the overall scalability separately. 

First, Figure~\ref{fig:breakdown} shows that the parallel sampling step is the dominant part on all evaluated graphs (takes an average of $83.30\%$ of the total time) and has a strong scalability due to the independent nature of each sampling operation;  \addd{thus, the scalability of {\hbmax} is affected by the encoding step and selection step.}   

Second, \add{for {\huffmax}, we note that building the Huffman codebook does not affect the overall scalability. This is because we only build up the codebook once during the warm-up phase, which takes an average of $2.19\%$ of the total time. Also, the Huffman encoding takes an average of $8.77\%$ of the total time as it only includes a small number of synchronizations to build frequency table. Thus, {\huffmax}'s encoding shows a strong scalability up to 64 cores.}

Third, \add{for all the large graphs (Pokec, LiveJournal, Arabic, Twitter7), the encoding of {\bitmax} shows a strong scalability up to 64 cores due to its independent nature. The bitmap encoding takes an average of $8.19\%$ of the total time.}

\add{Fourth, the selection operation in both {\huffmax} and {\bitmax} does not affect the overall scalability. 
First, our optimized reduction with parallel merge plays an important role, as shown in Figure \ref{fig:parallel_merge}, to minimize the  synchronization between threads. Without this optimization, the selection operation would become the bottleneck due to the original non-scalable parallel reduction.} \addd{Second, both the selection operations of {\huffmax}  and {\bitmax} are parallelized considering the NUMA effect so that each thread can maintain its local frequency table by only accessing the locally encoded data.} \add{Note that selection of {\bitmax} scales better than {\huffmax} because it does not have a decoding step.} 

Finally, we note that for the two small graphs DBLP and YouTube, \addd{Ripples does not scale well. This is because of the highly imbalanced workload between vertices in the seed-selection step, which can be also reflected in its high skewness score. While {\hbmax} achieves better scalability (can scale up to 32 cores) thanks to the proposed parallel merge and the consideration of NUMA effect.} The performance starts to degrade from 64 cores due to the relatively small workload. For the other six larger graphs, {\hbmax} can scale up to 64 cores and 
achieve an average speedup of 12.98$\times$ on 64 cores. 

\section{Conclusions and Future Work}\label{sec:Conclusion}
Graph analytics has emerged as an important class of data analytics tools. The ubiquity of data from a wide variety of sources has necessitated the development of scalable graph analytics tools. 
Influence Maximization (IM) 
is an important graph problem with many applications. Wide use of IM is currently limited due to the limitations from memory and computation requirements. 
In this paper, we presented \add{a ``compress-to-compute'' approach to use Huffman or bitmap coding to encode the sampled RRR sets. By exploiting the data locality 
and leveraging the efficient bit operations,} 
our method is able to reduce the peak memory usage by up to \add{$82.1\%$} and reduce computation by up to $15.0\%$ without noticeable loss of accuracy.  

Future research includes: 
i) establishing analytical bounds on the loss of information so that approximation guarantees of IM algorithms can be established;
ii) extension to distributed platforms with GPU accelerators, and 
iii) extension of compression techniques to other key graph algorithms (with large memory footprints) \addd{that address fundamental graph problems such as triangle counting, community detecting, and network alignment, 
which can potentially benefit from our proposed vertex-encodings}.


\section*{Acknowledgments}

\small 
The research is supported by the U.S. DOE ExaGraph project at Pacific Northwest National Laboratory (PNNL) and by the NSF awards OAC-2034169, OAC-2042084, OAC-1910213, and CCF-1815467 at Washington State University. 
This work used the Bridges-2 system, which is supported by the NSF award OAC-1928147, at Pittsburgh Supercomputing Center.


\newpage
\bibliographystyle{ACM-Reference-Format}
\bibliography{refs}


\end{document}